\documentclass[aps,nofootinbib,preprintnumbers,citeautoscript,twocolumn]{revtex4}
\usepackage{amsmath,amssymb}
\usepackage{xcolor}
\usepackage{enumerate}
\usepackage[colorlinks=true,citecolor=red,linkcolor=purple,linkbordercolor=cyan]{hyperref}

\usepackage{graphicx}
\graphicspath{{/home/kapil/My_data/kapilpaper/arXiv/Herring_Flicker_Distance-Paper/arxiv_Herring_Flicker/}}

\date{\today}

\begin{document}
\title{Herring-Flicker coupling and thermal quantum correlations in bipartite system}

\author{Kapil K. Sharma$^\ast$  \\
\textit{Department of Electrical Engineering,\\
Indian Institute of Technology Bombay, Mumbai 400076, India} \\
E-mail: $^\ast$iitbkapil@gmail.com}
\begin{abstract}
In this letter we study thermal quantum correlations as quantum discord and entanglement in bipartite system imposed by external magnetic field with Herring-Flicker coupling ie. $J(R)=1.642 e^{-2 R} R^{5/2}+O(R^{2}e^{-2R})$. The Herring-Flicker coupling strength is the function of $R$, which is the distance between spins and systems carry XXX Heisenberg interaction. By tuning the coupling distance $R$, temperature and magnetic field quantum correlations can be scaled in the bipartite system. We find the long sustainable behaviour of quantum discord in comparison to entanglement over the coupling distance $R$. We also investigate the situations, where entanglement totally dies but quantum discord exist in the system. The present findings in the letter may be useful for designing quantum wires, data bus, solid state gates and quantum processors.  
\end{abstract}
\maketitle


\section{Introduction}
Entanglement \cite{EPR} is the back bone for quantum information processing tasks, such as teleportation, quantum cryptography, quantum games and many others \cite{Nielsen_Chaung}. Fundamental role of entanglement in developing quantum computer has significant importance. As far as the quantum computer architecture is concerned, the quantum bus \cite{qb} have important role which is used to connect the quantum devices over the bus and routing the information. However there is no perfect model for quantum computer architecture as yet. Quantum bus based on spin chains \cite{Ird,SG} play the role to transport the data over the bus among quantum registers and the flow of data can be controlled by tuning the relative coupling among spins. So, in this direction it becomes very important to study the quantum correlations and their variations through coupling distance in spin chains. Varieties of spin chains have been studied by various authors without assuming the coupling strength as function of position in various configurations like XXX, XYZ, XXZ. Very few studies have been done for the effect of coupling distance on quantum correlations. In $1988$, Haldane and Shastry have studied the spin chains for long range interactions \cite{Hal,Sh}, in which the coupling strength follow the inverse square law. Relatively in the same direction, XXZ Heisenberg spin chain with long range interactions has been studied by BO. Li\cite{Bo}, XX Heisenberg spin chain with Calogero Moser type interaction has been studied by MA XiaoSan \cite{Ma}. Very recent studies for the same has been found in literature \cite{2017_long1,2017_long2,2017_long3,2017_long4}. In 2005, Zhen Huang and Sabre Kais have shown the dependency of entanglement on Herring-Flicker (HF) coupling distance \cite{HF} of XY spin chain governed by Ising modal \cite{Sabre}. The authors have found, the increasing amount of magnetic field decreases the entanglement over the HF coupling distance. HF coupling has its importance in determining the energy difference between triplet and singlet state of the Hydrogen molecule \cite{HF}, which is given by $J(R)=E_{triplet}-E_{singlet}=1.642 e^{-2 R} R^{5/2}+O(R^{2}e^{-2R})$. So by tuning the coupling distance $R$, the energy difference can be scaled. By taking the motivations from the above studies in continuation of Zhen Huang and Sabre Kais study \cite{Sabre}, we study the effect of HF coupling distance on thermal quantum correlations in XXX configuration of Hisenberg bipartite system and investigated that thermal quantum discord \cite{ds1,ds2,ds3} sustain over the larger range of $R$ in comparison to thermal entanglement \cite{TE1,TE2,TE3}. Further we find the parameter values of temperature and magnetic field over which entanglement vanish in the system but thermal quantum discord exists over $R$. Here we mention that quantum information processing community is always interested to search such systems which can persist long quantum correlations and avoid the phenomenon of entanglement sudden death \cite{ent1,ent2,ent3,ent4,ent5,ent6,ent7,ent8} for practical applicability of the system. Finding the properties of thermal quantum correlations through HF coupling distance can be useful to construct the quantum buses, solid state quantum gates and quantum processors. To the best of our knowledge, this study present the first outcome of thermal quantum discord over the HF coupling distance $R$.

The Plan of the paper is as follows in section $2$, we discusses the Hamiltonian, thermal density matrix and concurrence. In  Section $3$, we study the dynamics of thermal quantum discord and entanglement with the varying values of parameters temperature and magnetic field over the coupling distance $R$. In the last with Section $4$, we present the conclusion of the paper.
\section{Hamiltonian, thermal density matrix, concurrence and discord}
In this section we give the Hamiltonian of the bipartite spin system, thermal density matrix and concurrence. The Hamiltonian of bipartite system is given as below,
\begin{equation}
H=J_{x}\sigma_{1}^{x}\sigma_{2}^{x}+J_{y}\sigma_{1}^{y}\sigma_{2}^{y}+J_{z}\sigma_{1}^{z}\sigma_{2}^{z}
+B(\sigma_{1}^{z}+\sigma_{2}^{z})
\end{equation}
In this letter we consider the case with $(J_{x}=J_{y}=J_{z}=J>0)$ for XXX configuration of Heisenberg spin chain. Further we assume the coupling $J$ is HF coupling ie.$J(R)$, given as,
\begin{equation}
J(R)=1.642 e^{-2 R} R^{5/2}+O(R^{2}e^{-2R})
\end{equation}
So the modified Hamiltonian for XXX configuration with HF coupling can be obtained as below,
\begin{equation}
H=J(R)[\sigma_{1}^{x}\sigma_{2}^{x}+\sigma_{1}^{y}\sigma_{2}^{y}+\sigma_{1}^{z}\sigma_{2}^{z}+B(\sigma_{1}^{z}+\sigma_{2}^{z}]
\label{hr}
\end{equation} 
\begin{figure}
\centering
     \includegraphics[width=0.48\textwidth]{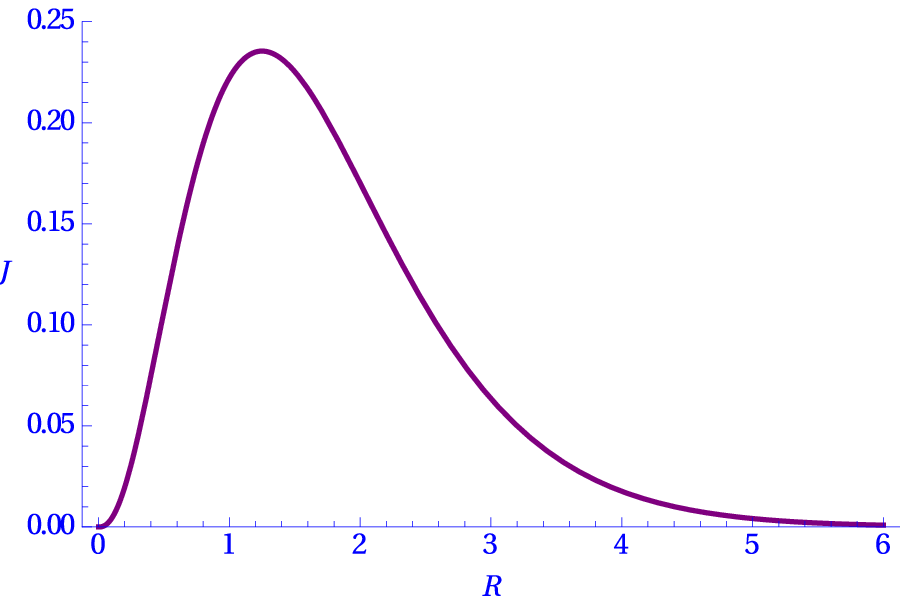}
       \caption{Herring-Flicker copuling strength $J$ w.r.t R}\label{hf1}
\end{figure}
The plot of the function $J(R)$ is shown in the Figure \ref{hf1}.
In order to study the thermal behaviour of entanglement we need to calculate the thermal density matrix of the system, which can be obtained as,
\begin{equation}
\rho_{T}=\frac{e^{-H/KT}}{Tr(e^{-H/KT})}
\end{equation}
The matrix form of $\rho^{T}$ can be obtained by calculating the eigenspectrum of the Hamiltonian given in Eq.\ref{hr}, this matrix is obtained below,
\begin{equation}
\rho_{T}=[c_{1},c_{2},c_{3},c_{4}] \label{tm}
\end{equation}
with,
\begin{eqnarray}
c_{1}=[a_{11},0,0,0]^{T}, \quad c_{2}=[0,a_{22},a_{32},0]^{T} \\
c_{3}=[0,a_{23},a_{33},0]^{T}, \quad c_{4}=[0,0,0,a_{44}]^{T}
\end{eqnarray}
with,
\begin{widetext}
\begin{eqnarray}
a_{11}=\frac{e^{-\frac{2 B+J(R)}{2 \text{KT}}}}{e^{-\frac{J(R)-2 B}{2 \text{KT}}}+e^{-\frac{2 B+J(R)}{2 \text{KT}}}+2 e^{-\frac{J(R)}{2 \text{KT}}}+2 e^{\frac{3J}{2\text{KT}}}},
 a_{22}=\frac{e^{-\frac{J(R)}{2 \text{KT}}}+e^{\frac{3 J(R)}{2 \text{KT}}}}{e^{-\frac{J(R)-2 B}{2 \text{KT}}}+e^{-\frac{2 B+J(R)}{2 \text{KT}}}+2 e^{-\frac{J(R)}{2\text{KT}}}+2 e^{\frac{3J(R)}{2 \text{KT}}}},\\ \\
a_{23}=\frac{e^{-\frac{J(R)}{2 \text{KT}}}-e^{\frac{3 J(R)}{2 \text{KT}}}}{e^{-\frac{J(R)-2 B}{2 \text{KT}}}+e^{-\frac{2 B+J(R)}{2 \text{KT}}}+2 e^{-\frac{J(R)}{2\text{KT}}}+2 e^{\frac{3J(R)}{2 \text{KT}}}}, \quad
a_{32}=\frac{e^{-\frac{J(R)}{2 \text{KT}}}-e^{\frac{3 J(R)}{2 \text{KT}}}}{e^{-\frac{J(R)-2 B}{2 \text{KT}}}+e^{-\frac{2 B+J(R)}{2 \text{KT}}}+2 e^{-\frac{J(R)}{2\text{KT}}}+2 e^{\frac{3 J(R)}{2 \text{KT}}}}\\ \\
a_{33}=\frac{e^{-\frac{J(R)}{2 \text{KT}}}+e^{\frac{3 J(R)}{2 \text{KT}}}}{e^{-\frac{J(R)-2 B}{2 \text{KT}}}+e^{-\frac{2 B+J(R)}{2 \text{KT}}}+2 e^{-\frac{J(R)}{2\text{KT}}}+2 e^{\frac{3 J(R)}{2 \text{KT}}}}, \quad
a_{44}=\frac{e^{-\frac{J(R)-2 B}{2 \text{KT}}}}{e^{-\frac{J(R)-2 B}{2 \text{KT}}}+e^{-\frac{2 B+J(R)}{2 \text{KT}}}+2 e^{-\frac{J(R)}{2 \text{KT}}}+2 e^{\frac{3 J(R)}{2\text{KT}}}}
\end{eqnarray}
\end{widetext}
Further the format of thermal density matrix gives the clue to obtain the concurrence in $4\times 4$ dimensional matrix. Here we mention that concurrence is a good measure of entanglement for bipartite system. The concurrence is $4\times 4$ dimensional matrix is given by
\begin{figure*}
\centering
     \includegraphics[width=0.9\textwidth]{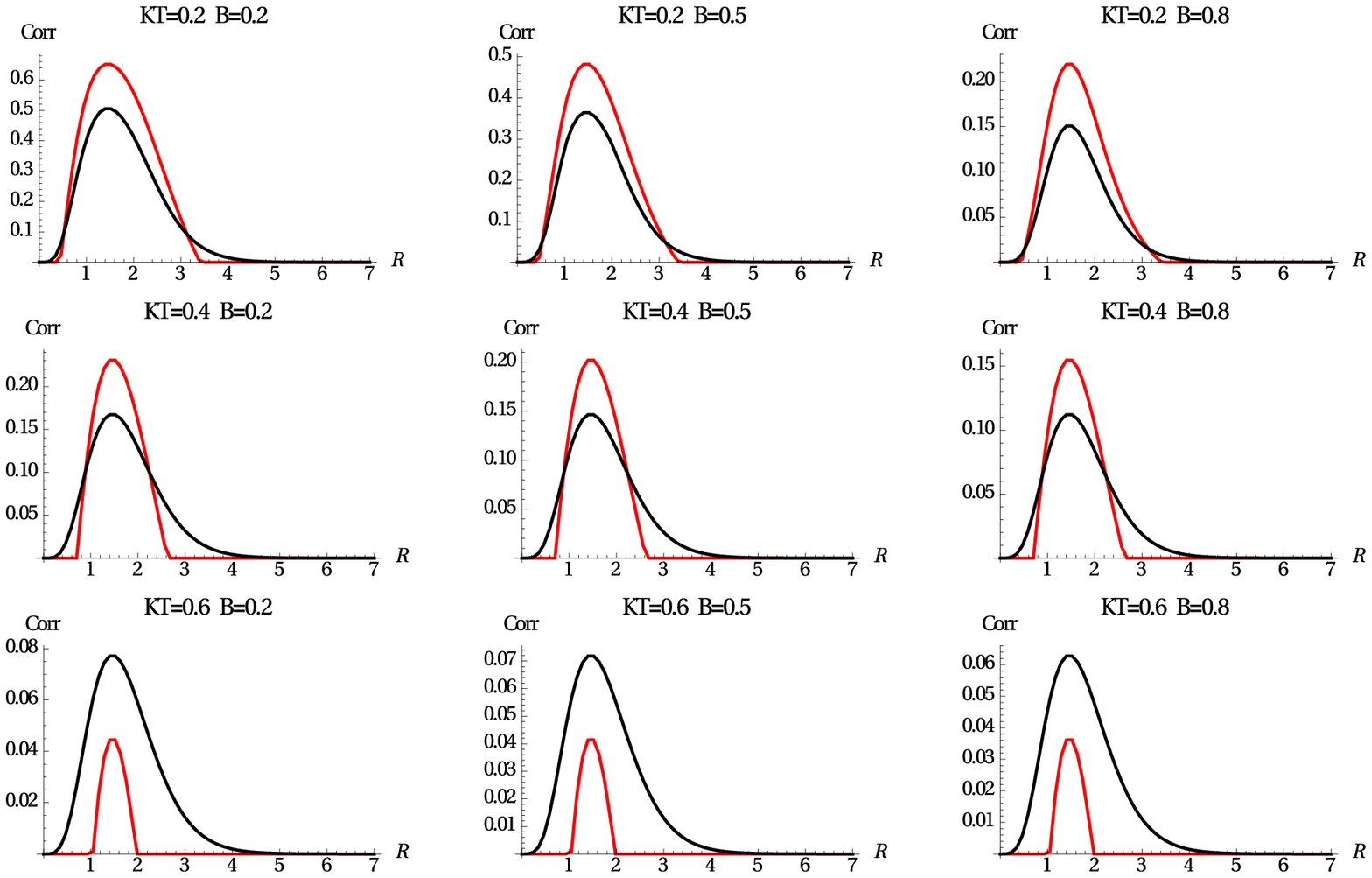}
       \caption{Plot of entanglement and quantum discord over the HF coupling distance $R$. Here red color (\text{\color{red}$\curlywedge$}) is entanglement and black color (\text{\color{black}$\curlywedge$}) is quantum dicord}\label{qcf}
\end{figure*}

\begin{figure*}
\centering
     \includegraphics[width=0.9\textwidth]{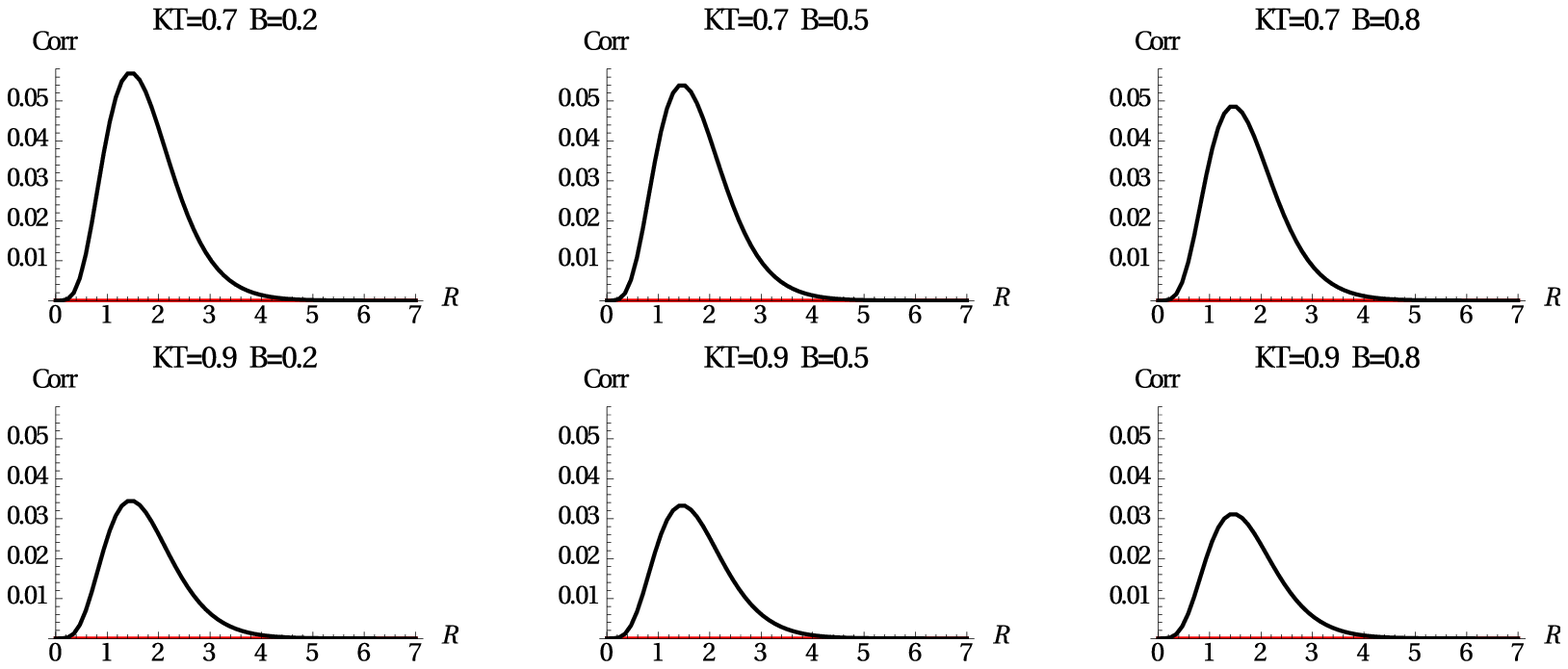}
       \caption{Plot of entanglement and quantum discord over the HF coupling distance $R$. Here red color (\text{\color{red}$\curlywedge$}) is entanglement and black color (\text{\color{black}$\curlywedge$}) is quantum dicord}\label{qdf}
\end{figure*}

\begin{equation}
C(\rho)=max\{0,p-q-r-s\}
\end{equation}
where $(p>q>r>s)$ with $(p=\sqrt{\lambda_{1}},q=\sqrt{\lambda_{2}},r=\sqrt{\lambda_{3}},s=\sqrt{\lambda_{4}})$. Here $(\lambda_{1},\lambda_{2},\lambda_{3},\lambda_{4})$ are the eigenvalues of the matrix $\rho \rho^{f}$. Where $\rho^{f}$ is the spin flip matrix given as,
\begin{equation}
\rho^{f}=(\sigma_{y}\otimes \sigma_{y}) \rho^{*} (\sigma_{y}\otimes \sigma_{y})
\end{equation}
Here $\rho^{*}$ is the complex conjugate of the density matrix.
Here we calculate the concurrence in the thermal density matrix given in Eq. \ref{tm}. In this case the concurrence is the function of three parameters $(KT,B,R)$.

Here we mention another quantum correlation so called quantum discord. The quantum discord is measurement based quantum correlation, for bipartite system $AB$, it is mathematically defined as below,
\begin{equation}
Q(\rho^{AB})=S(\rho^{A})-S(\rho^{AB})+\min_{\{{A_{k}}\}}S(\rho^{B}|A_{k}).
\end{equation}
Where $S(\rho^{AB})$ is the Von Neumann entropy in bipartite density matrix, $S(\rho^{A})$ is Von Neumann entropy in marginal density matrix $\rho^{A}$ and the term $\min_{\{{A_{k}}\}}S(\rho^{B}|A_{k}$ is the minimization of Von Neumann entropy in marginal density matrix $\rho^{B}$ when the POVM $\{A_{k}\}$ is performed on subsystem $A$. Fore more details and Mathematica code to calculate quantum discord, one can follow Ref\cite{ds1,ds2,ds3,ent5}.

\section{Thermal quantum correlations over HF coupling}
In this section we present the dynamics of thermal quantum correlations (quantum discord, entanglement), over the HF coupling distance $R$ with the varying parameters of temperature and magnetic field. The results have shown in Fig.\ref{qcf}. For a fixed value of parameter $KT=0.2$, as the value of magnetic field $B$ increases, the amplitude of entanglement and quantum discord decreases. It is observed that as HF coupling distance advances, both entanglement and quantum discord slowly vanish after a certain range of $R$. It is interesting to observe that quantum discord sustain over the large range of HF coupling distance $R$, while entanglement dies at $R=3.2$. As the value of KT increases with $(KT=0.4)$, the amplitude of both quantum correlations decrease and sustainable range of entanglement also decrease, however quantum discord sustain over the large range of $R$  and slowly surpass over the entanglement. Further with $KT=0.6$, we again find the amplitude of quantum correlations decrease, and sustainable range of entanglement also decrease than previous cases. With the same value $KT=0.6$, quantum discord completely cover the entanglement and even present in the absence of entanglement in the system. Here we mention that quantum discord is strong quantum correlation and has the long range sustain ability over the entanglement. Next we find with the parameter value $KT\geq 0.7$ and $\forall B$, the entanglement totally vanish in the system but quantum discord still exist. These results have been shown in Fig.\ref{qdf}. The value $(KT\geq 0.7)$ is  the threshold value at which the drastic change takes place in the system.

\section{Conclusion}
In this present article, we have studied thermal quantum correlations in bipartite system with the existence of Herring-Flicker coupling. The system carry XXX Heisenberg interaction and imposed by external magnetic field. We have found for the fixed value of temperatures with increasing amount of magnetic field, the amplitude of quantum correlations decrease. Further it is observed that quantum discord sustain over the large range of HF coupling distance $R$ in comparison to entanglement. The increasing values of the temperature decrease the amplitude of quantum correlations and increasing values of magnetic field effect the sustainable range of entanglement. While sustainable range of quantum discord has less influenced by increasing magnetic field. At the parameter value of temperature $(KT=0.6)$, quantum discord become dominant over the entanglement and totally surpass it. We also have found that, increasing values of the parameters with $(KT\geq 0.7)$ and $\forall B$, kill the entanglement, while quantum discord still exist in the system. So quantum discord is more robust quantum correlation in comparison to entanglement, which is generally predicted behaviour of quantum discord. We have found long range sustain ability of the quantum discord over the entanglement as the coupling distance increases with increasing amount of magnetic field. The present study can be improved in another spin chain configurations like XXZ, XYZ, XXZ etc. to know the behaviour of quantum discord and entanglement with Herring-Coupling, and can be useful for quantum information processing.

\section{Acknowledgement}
The authors acknowledge support from the Ministry of Electronics \& Information Technology, Government of India, through the Centre of Excellence in Nano-Electronics, IIT Bombay.

\end{document}